\documentclass{iopart}
\usepackage{graphics}
\begin{document}
\bibliographystyle{apsrev}

\title[Classical Wave Delays in Quantum Scattering]{Manifestation of 
classical wave delays in a fully-quantized model of the scattering 
of a single photon}

\author{Thomas Purdy\dag\footnote[3]{Current address: Department of 
Physics, University of California, Berkeley, CA 94720-7300, USA},
Daniel R Taylor\ddag\footnote[4]{Current address: Department of Physics, 
Boston University, Boston, MA 02215, USA} and 
Martin Ligare\ddag\footnote[5]{To whom correspondence should be 
addressed (mligare@bucknell.edu)}}

\address{\dag\  Department of Physics, Carnegie Mellon 
University, Pittsburgh, PA 15213, USA}

\address{\ddag\ Department of Physics, Bucknell University, Lewisburg, 
PA 17837, USA}

\begin{abstract}
We consider a fully-quantized model of spontaneous emission,
scattering, and absorption, and study propagation of a single photon
from an emitting atom to a detector atom both with and without an
intervening scatterer.  We find an exact quantum analog to the
classical complex analytic signal describing an electromagnetic wave
scattered by a medium of charged oscillators.  This quantum signal
exhibits classical phase delays.  We define a time of detection which,
in the appropriate limits, exactly matches the predictions of a
classically defined delay for light propagating through a medium of
charged oscillators.  The fully quantized model provides a simple,
unambiguous, and causal interpretation of delays that seemingly imply
speeds greater than $c$ in the region of anomalous dispersion.
\end{abstract}

\pacs{32.80.-t, 42.25.Bs, 42.50.Ct}



\section{Introduction}

The subtleties of appropriately defining the speed of light pulses in
dispersive media have been investigated for over a century.  Although
a careful classical analysis was performed early in the twentieth
century (see, for example, the papers collected in reference
\cite{BRI60}), recent observations of anomalously slow
\cite{HAU99,KAS99} and anomalously fast \cite{STE93,WAN00} speeds of
light have returned studies of light propagation to the pages of
contemporary physics journals, and a recent review so-called 
superluminal propagation is contained in \cite{CHI97}.  
In this paper we study the scattering
of single spontaneously emitted photons and develop a simple
fully-quantized microscopic model of light propagation in a dielectric
medium.  We also compare this quantized model to a simple microscopic
classical model of scattering from charged oscillators.  Although our
models are limited to the simplest case of weak scattering in linear
dielectrics, they serve as a basis for understanding propagation in
more complex media at the quantum level.

Classical pulses can be understood as the superposition of the
incident field with fields re-radiated by the atoms of the medium.
This point of view has been articulated clearly by Feynman
\cite{FEY63}, with recent elaborations by Sherwood \cite{SHE96} and
Milonni \cite{MIL96}.  The re-radiated fields are phase-shifted and
attenuated relative to the incident field, and the leading edges of
all fields propagate at the vacuum speed of light $c$.  We identify
quantum mechanical quantities that have a striking quantitative
parallel in the complex analytic signal describing fields in the
classical model.  Although the state vector describing the system has
no absolute overall phase, the scattering-induced phase shift of the
classical field is manifested in the description of the quantum field
as well as in the probability amplitude for atoms subsequently excited
by the field. The pulse re-shaping of classical fields that leads to
group delays is also evident in our quantum model, and results in
delays that are equivalent in the classical and quantum models.  All
delays are clearly due to the superposition of quantum effects that
propagate at the vacuum speed of light $c$, in spite of the appearance
of delays that seemingly imply so-called superluminal velocities.  The
parallels between classical and quantum scattered fields facilitate
the understanding of delays of single photons in terms of classical
concepts.

The classical medium we consider consists of charged point particles
attached to their equilibrium positions by linear restoring forces.
This model is considered in undergraduate textbooks
\cite{FEY63,GRI99}.  The classical medium has an index of refraction
which for a dilute collection of scatterers depends linearly on the
density of the scatterers, and the index of refraction determines the
phase and group velocities of radiation in the medium.

Our quantum system consists of a set of two-level atoms at fixed
positions interacting with the quantized modes of a one-dimensional
multimode optical cavity.  An initially excited atom spontaneously emits
radiation into the modes of the cavity (a photon); the radiation is
scattered by a second atom; and a third atom serves as a detector.  We
find an analytical expression describing the excitation of the
detector atom, and an expression for the expectation value of the
space- and time-dependent electromagnetic field intensity.  We then
identify an exact quantum analog to the complex analytic signal of
a classical field.  The scattering-induced time delays in the
quantum mechanical quantities thus have direct parallels in classical
delays.  We also find a simple, unambiguous, and causal interpretation
at the quantum level of a superluminal velocity that occurs in the
region of anomalous dispersion very near to an atomic resonance.

We note that the results obtained analytically in this paper may also
be observed using straightforward numerical techniques like those used
in references \cite{TAY01,LIG01a,LIG02,BUZ99}.  Our work bears a
particularly close relationship to the numerical work reported in
\cite{DRO00}, in which quantum interference effects are observed in
the scattering of single-photon wavepackets.

We review the classical model in section \ref{sec:classical}, and in
section \ref{sec:quantum} we discuss our quantum mechanical model and
present our analytical solution.  (We leave the details of the
derivation to the Appendix.)  We highlight the similarity of the
quantum and classical dynamics in section \ref{sec:comparison}, and
in section \ref{sec:cog:delay} we investigate a delay that seemingly
implies superluminal propagation in both the classical and quantum
models.

\section{Review of Scattering from Classical Oscillators}
\label{sec:classical}

A simple classical model of the interaction of radiation with matter
consists of an electromagnetic plane wave driving electrons attached to
molecules by linear springs \cite{FEY63,GRI99}.  This simple model is
able to account for the attenuation and phase shift of waves
transmitted through a dilute material comprised of such oscillators.

For a plane wave normally incident on a dielectric slab with thickness
$\Delta z$ and real index of refraction $n$, the delay in the arrival
of a point of constant phase on the far side of the medium (compared
to a wave traveling in vacuum) is determined by the phase
velocity $v_{\phi}=\omega/k = c/n$.  This delay is given by
\begin{equation}
\Delta t_{\phi}= \frac{\Delta z}{v_{\phi}} - \frac{\Delta z}{c}
	       = \frac{\Delta z}{c}(n-1).	
\label{eq:ph:delay}
\end{equation}
The delay in the arrival of the peak of a modulation envelope of a
quasi-monochromatic pulse is determined by the group velocity $v_{\rm
g}= \rmd\omega/{\rmd k}=c/(n+\omega\frac{\rmd n}{\rmd\omega})$, and is
given by
\begin{equation} 
\Delta t_{\rm g}= \frac{\Delta z}{v_{\rm g}}-\frac{\Delta z}{c}
	        = \frac{\Delta z}{c}\left(n-1 + 
                   \omega \frac{\rmd n}{\rmd \omega}\right).
\label{eq:g:delay}
\end{equation} 

For pulses that are not sufficiently monochromatic, the
simple concepts of phase and group velocity are inadequate to
characterize all of the effects of pulse-reshaping as the field
propagates.  Several other velocities and delays have been developed
(see, for example, references \cite{SMI70,BLO77}) and in this paper we
focus on a delay determined by the ``temporal center of gravity'' of
the field intensity of a pulse at a fixed position $z$ ``downstream''
from the dielectric, i.e.,
\begin{equation}
\Delta t_{{\cal E}^2} \equiv \left(\frac{\int t {\cal E}(z,t)^2\, \rmd t}
			           {\int  {\cal E}(z,t)^2\, \rmd t}
                                   \right)_{\mbox{with dielectric}} 
			- \left(\frac{\int t {\cal E}(z,t)^2\, \rmd t}
			             {\int  {\cal E}(z,t)^2\, \rmd t}
		                    \right)_{\mbox{no dielectric}}.  
\label{eq:cl:tcom:delay}
\end{equation}
This is closely related to concepts used to define the centrovelocity
in \cite{SMI70}.  For quasi-monochromatic pulses far from resonance
this delay is equivalent to the group delay, but in general it is
necessary to calculate explicitly the field $\cal{E}$ in order to
determine $\Delta t_{{\cal E}^2}$.  For the spontaneously emitted
pulses with Lorentzian spectrums that are considered in this paper we
will show that the ``temporal-center-of-gravity'' delay happens to be
equal to twice the group delay.

In order to find classical expressions for the index of refraction and
absorption we follow the development outlined by Feynman \cite{FEY63}.
We consider a monochromatic linearly-polarized plane wave of frequency
$\omega$ normally incident on an infinite thin slab of material with
density of oscillators $N$ and thickness $\Delta z$.  The plane of the
slab is normal to the $z$ direction, and located at $z=0$.  The
oscillators have natural frequency $\omega_0$, mass $m$, instantaneous
speed $v$, and are assumed to experience a damping force proportional
to the first derivative of acceleration, which for sinusoidal
oscillations is equivalent to a damping proportional to $-v$.  The
field on the far side of the slab is the superposition of the incident
field and the field scattered by the material.  In the steady state
the scattered field is proportional to the incident field, with an
amplitude and phase shift given by simple resonance theory. The
infinite slab geometry considered here creates an effectively
one-dimensional model that matches the simple quantum mechanical
system we introduce in the following section.

If the incident field is $E_{\rm i}=E_0\exp\left[-\rmi\omega(t-z/c)\right]$,
then the scattered field is \cite{FEY63,GRI99}
\begin{eqnarray}
E_{\rm s} &= -\frac{qN\Delta z}{2\epsilon_0c}v \nonumber \\ 
	  &= \rmi f \frac{\gamma\omega}
	    {(\omega_0^2-\omega^2 - \rmi\gamma\omega)}E_{\rm i},
\label{eq:e:sc}
\end{eqnarray}
where in the last line we have introduced the dimensionless parameter
$f=q^2N\Delta z/(2m\epsilon_0c\gamma)$ which characterizes the
magnitude of the scattering..  The total transmitted field $E_{\rm t}$
on the far side of the material is
\begin{eqnarray}
E_{\rm t} &= E_{\rm i} + E_{\rm s} \nonumber \\
	  &= E_{\rm i}\left[1 + \rmi f\left(\frac{\omega\gamma}
                 {\omega_0^2-\omega^2-\rmi\gamma\omega}\right)\right].
\label{eq:e:tr}
\end{eqnarray}

In comparing classical predictions with those from a quantum
mechanical model we will be interested in the limit in which the
detuning from resonance is small compared to the natural frequency of
the oscillator, i.e., $\vert \omega - \omega_0\vert \ll \omega_0$.
Taking advantage of this condition we rewrite the transmitted field of
equation (\ref{eq:e:tr}) in terms of the detuning from resonance
\begin{equation}
\delta = \omega-\omega_0,
\label{eq:cl:approx}
\end{equation}
and we approximate other occurrences of $\omega$ with $\omega_0$, giving 
\begin{equation}
E_{\rm t} \simeq E_{\rm i}\left[1 - \rmi f\left(\frac{\gamma}
	         {2\delta + \rmi\gamma}\right)\right]. 
\label{eq:e:tr:approx}
\end{equation}

For weak scattering, i.e., when $f \ll 1$, the transmitted field is
approximately
\begin{eqnarray}
E_{\rm t} &\simeq E_{\rm i}\left[1 - f\frac{\gamma^2}
	                 {\left(4\delta^2 + \gamma^2\right)}\right]
	         \left[1- \rmi f \frac{\gamma\delta}
                {\left(4\delta^2 + \gamma^2\right)}
		\right]    \nonumber \\
           &\simeq E_{\rm i} \exp\left[-f\frac{\gamma^2}
	             {\left(4\delta^2 + \gamma^2\right)}\right]
	        \exp\left[-\rmi f\frac{\gamma\delta}
                         {\left(4\delta^2 + \gamma^2\right)}\right].  
\label{eq:e:trans}
\end{eqnarray}
The phase-shift and attenuation given by equation (\ref{eq:e:trans})
lead to the index of refraction
\begin{equation}
n = 1 - \frac{Nq^2}{m\epsilon_0\omega_0\gamma}\frac{\gamma\delta}
			{4\delta^2 + \gamma^2}
\label{eq:index}
\end{equation}
and intensity absorption coefficient
\begin{equation}
\alpha =  \frac{Nq^2}{m\epsilon_0c\gamma}\frac{\gamma^2}
{(4\delta^2 + \gamma^2)}.
\label{eq:alpha}
\end{equation}

\begin{figure}
\begin{center}
\includegraphics{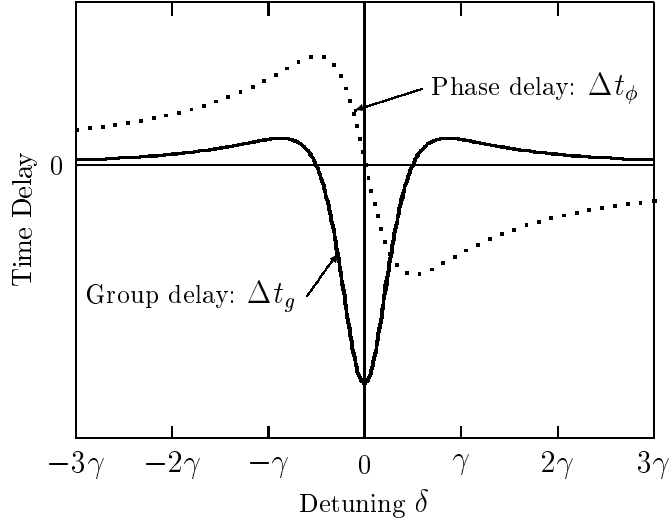}
\end{center}
\caption{Phase and group delays of a plane wave propagating
through a medium of classical oscillators.  The group delay is 
negative in the region of anomalous dispersion, i.e., for
$-\gamma/2<\delta <\gamma/2$.}
\label{f:delay}
\end{figure}
Using the expression for the index of refraction given in equation
(\ref{eq:index}), and assuming $\delta\ll \omega_0$, the phase and
group delays can be written
\begin{equation}
\Delta t_{\phi} \simeq -\frac{2f}{\omega_0}
		      \left(\frac{\gamma\delta}{4\delta^2 + \gamma^2}\right) 
\label{eq:phase:delay}, 
\end{equation}
\begin{equation}
\Delta t_{\rm g} \simeq 2f\gamma\frac{(4\delta^2 - \gamma^2)}
                                    {(4\delta^2 + \gamma^2)^2}.
\label{eq:group:delay}
\end{equation}

The phase and group delays predicted from the classical model are
illustrated in figure \ref{f:delay}.  The group delay is positive away
from resonance, indicating a group velocity less than the vacuum speed
of light $c$.  In the region of anomalous dispersion near resonance
the classical group delay is negative, indicating that the group
velocity is greater than $c$.  The steep slope of the index of
refraction and the accompanying strong and rapidly varying absorption
severely distort pulses that are not sufficiently monochromatic, and
the standard physical interpretation of $\rmd\omega/\rmd k$ as the
speed at which the peak of a modulation envelope travels is not valid
for such pulses in this region.  The ``temporal-center-of-gravity''
delay retains obvious physical meaning for all pulses, and in section
\ref{sec:cog:delay} we develop quantum mechanical properties that are
analogous to this delay for spontaneously emitted photons. This
provides a framework for an unambiguous and causal interpretation of
negative values of delays in both classical and quantum models. 

Group delays and ``temporal-center-of-gravity'' delays can be
understood as a result of the transient oscillations in the medium
\cite{FEY63}.  These transient oscillations radiate fields that
initially cancel the field of the incident wave, modifying the leading
edge of transmitted pulses before they settle down to the steady-state
fields given by equation (\ref{eq:e:trans}).  Field transients must be
included explicitly in the calculation of delays for pulses in the
region of anomalous dispersion.  We include a specific calculation of
the ``temporal-center-of-gravity'' delay in section
\ref{sec:cog:delay}.

\section{Quantum Model}
\label{sec:quantum}

\begin{figure}
\begin{center}
\includegraphics{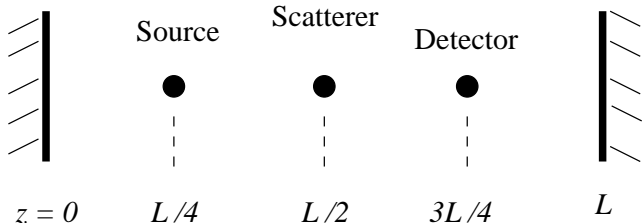}
\end{center}
\caption{Quantum mechanical model consisting of two-level atoms at 
fixed positions in a one-dimensional multimode optical cavity.}
\label{f:model}
\end{figure}

The quantum mechanical system we consider is illustrated in
figure \ref{f:model}, 
and consists of three two-level atoms in a one-dimensional multimode
optical cavity that extends from $z=0$ to $z=L$.  The leftmost atom
will initially be in the excited state, and will be the source of the
field.  The atom in the middle will scatter the radiation emitted by
the source atom, and the atom on the right will serve as the detector.
The cavity is assumed to be large in the sense that the length $L$ is
very much greater than the wavelength of the light emitted by the
atoms.  (The finite optical cavity does not contribute to the physical
phenomena under investigation; it simply provides a convenient
quantization volume for the field modes used in our calculations.)  In
this section we discuss the standard quantum optical Hamiltonian that
we use, and we present the analytical solution for the time dependence
of this system.

The zero-field resonance frequencies of the atoms are labeled
$\omega^{\rm (at)}_j$, where $j=1$, 2, or 3, and the positions of the
atoms will be labeled $z_j$.  In the results given below we will
assume that the atoms are at positions $z_1=L/4$, $z_2=L/2$, and
$z_3=3L/4$, as illustrated in figure \ref{f:model}, although the delay
times we derive do not depend on the exact positions.  The
standing-wave field modes of the cavity are separated in angular
frequency by the fundamental frequency
\begin{equation}
\Delta_{\rm c} = \pi\frac{c}{L}.
\end{equation}
This mode spacing may be small enough that many modes fall within the
natural line-width of the atoms.  

For convenience we assume that the frequency of one of the modes
corresponds exactly to the resonance frequency of atom 1, the emitting
atom, and that the length of the cavity is such that it contains an
even number of wavelengths of this mode. We label the frequency of
this mode $\omega_0=m_0\Delta_{\rm c}$, where $m_0$ is an integer
divisible by 4.  (This assumption affects the details of some of our
calculations, but not the existence of an analytic solution, nor any
of our results concerning delay times.  Classically, this assumption
assures that optical path length between any pair of atoms is an
integer number of half-wavelengths of the resonant radiation from atom
1.)  The other mode frequencies will be enumerated from this mode so
that
\begin{equation}
\omega_m= (m_0+m)\Delta_{\rm c},
\label{eq:enumerate}
\end{equation}
where $m=0,\pm 1,\pm 2,\dots$ Our large-cavity limit also assures that
the frequencies of the three atoms are all very much greater than the
fundamental frequency of the cavity, i.e., $\omega^{\rm (at)}_j\gg
\Delta_{\rm c}$.  In this limit the atoms interact with a large number
of modes, and not simply the resonant mode.

As in the classical case we wish to study the effects of the detuning
of the source field on the scattering of the radiation.  We use the
same symbol $\delta$ as in the classical case to represent the
detuning of the field, but in the quantum case the detuning is
directly tied to the properties of the source and scattering atoms:
\begin{equation}
\delta = \omega_1^{\rm (at)} - \omega_2^{\rm (at)}.
\end{equation}
The detector atom (atom 3) is assumed to have the same resonance
frequency as the source atom, i.e., $\omega^{\rm (at)}_1 = \omega^{\rm
(at)}_3$.

We use as basis states the eigenstates of the atomic plus free-field
Hamiltonian
\begin{eqnarray}
\hat{H}_0 &= \hat{H}_{\rm atoms} + \hat{H}_{\rm field} \nonumber \\
	  &= \sum_{j=1}^3\hbar\omega^{\rm (at)}_j\sigma^z_j + \sum_m
	      \hbar\omega_m a^\dagger_m a_m,
\end{eqnarray}
where $\sigma^z_j$ is the third component of the atomic pseudo-spin
operator of atom $j$, and $a_m$ and $a^\dagger_m$ are the lowering and
raising operators for the $m^{\rm th}$ field mode.  (We have re-zeroed
the energy scale to remove zero-point energy of the field modes.) The
basis states will be denoted as follows:
\begin{itemize}
\item $\vert e,g,g;0\rangle$ --- Atom 1 excited, atoms 2 and 3 in ground 
state; no photons in field,
\item $\vert g,e,g;0\rangle$ --- Atom 2 excited, atoms 1 and 3 in 
ground state; no photons in field,
\item $\vert g,g,e;0\rangle$ --- Atom 3 excited, atoms 1 and 2 in 
ground state; no photons in  field,
\item $\vert g,g,g;1_m\rangle$ --- All atoms in ground state; one photon
in field mode with frequency $(m_0 + m)\Delta_{\rm c}$. 
\end{itemize}

We use the standard electric-dipole and rotating-wave approximations
in the interaction Hamiltonian \cite{MEY99,SAR74, LOU83} to give
\begin{eqnarray}
\hat{H} &=& \hat{H}_{\rm atoms} + \hat{H}_{\rm field} + 
                        \hat{H}_{\rm interaction}  \nonumber\\
        &=& \hat{H}_0 +
	    \sum_{j=1}^3\sum_m \hbar\left(g_{jm}a_m \sigma_j^+ 
			             + g_{jm}^\ast a_m^\dagger \sigma_j^-
						\right),
\label{eq:hamiltonian}
\end{eqnarray}
where the strength of the coupling of the $j^{\rm th}$ atom to the
$m^{\rm th}$ mode of the field is characterized by the constant
$g_{jm}$, and $\sigma_j^+$ and $\sigma_j^-$ act as raising and
lowering operators for atom $j$.

In our large-cavity limit, $\omega^{\rm (at)}_j\gg \Delta_{\rm c}$,
we can make the approximation that all modes that influence the
dynamics of the system are near the atomic resonances, and the
atom-field coupling constants are given by
\begin{equation}
g_{jm}= \Omega_j\sin\left[(m_0+m)\pi z_j/L\right].
\end{equation}
In the previous equation $\Omega_j$ is a constant given by 
\begin{equation}
\Omega_j = d_j\left(\frac{\omega^{\rm (at)}_j}{2\hbar\epsilon_0 V}
			\right)^{1/2},
\end{equation} 
where $d_j$ is the dipole matrix element between the levels of atom
$j$, and  $V$ is the effective volume of the cavity.

Initially only the source atom will be excited and no photons will be
present in the field, so that
\begin{equation}
\vert\psi(0)\rangle = \vert e,g,g;0\rangle,
\end{equation}
and we write the general state of the system  as the linear combination
\begin{equation}
\fl \vert\psi(t)\rangle =  c_1(t)\vert e,g,g;0\rangle 
		         + c_2(t)\vert g,e,g;0\rangle 
		         + c_3(t)\vert g,g,e;0\rangle   
		         + \sum_m b_m(t)\vert g,g,g;1_m\rangle.
\label{eq:gen:psi}
\end{equation}
The Schr\"{o}dinger equation yields the following set of coupled
differential equations for the coefficients in equation (\ref{eq:gen:psi}):
\begin{equation}
\dot{c}_j = -\rmi\left(\omega^{\rm (at)}_j c_j + \sum_m g_{jm}b_m\right) 
\label{eq:c:schro}
\end{equation}
\begin{equation}
\dot{b}_m = -\rmi\left(\omega_m b_m +  g_{1m}^\ast c_1 + g_{2m}^\ast c_2 + 
		g_{3m}^\ast c_3 \right).
\label{eq:b:schro}
\end{equation}

We solve this set of equations with the Laplace transform technique
used by Stey and Gibberd \cite{STE72}.  Laplace transforms have also
been used to solve the Schr\"{o}dinger equation in similar problems
with two interacting atoms in three-dimensions \cite{MIL74,MIL75}.
Because the Laplace transform technique is not new, and because we
would like to focus on analogies with the classical model and physical
interpretation, we leave the details of our solution to the appendix,
and simply quote our results here.

\begin{figure}
\begin{center}
\includegraphics{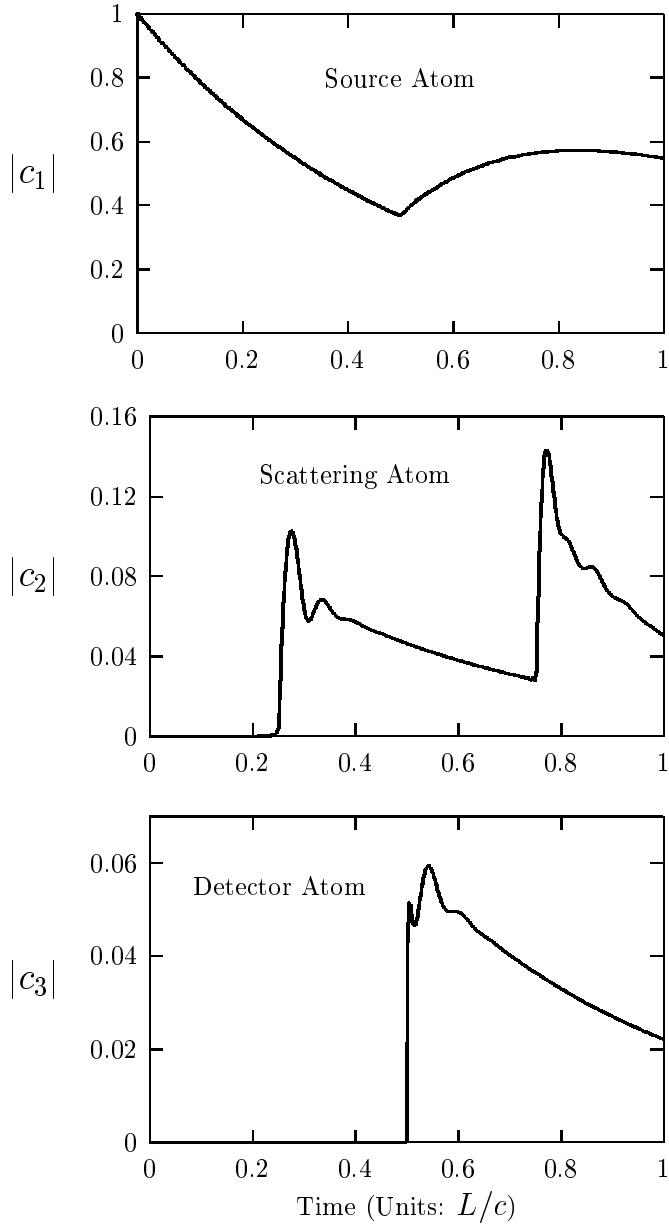}
\end{center}
\caption{Magnitude of the amplitudes for the atoms to be in the
excited state, starting from the state $\vert\psi(0)\rangle = \vert
e,g,g;0\rangle$. The decay rates of the atoms are $\gamma_1 =
4$, $\gamma_2 = 64$, and $\gamma_3 = 1024$ (in the units of the figure)
and the detuning is $\delta = 1.56\gamma_2$.}
\label{f:sol2}
\end{figure}

The general features of the solution giving the time-dependent atomic
excitation amplitudes are illustrated in figure \ref{f:sol2}.  The
initially excited atom decays exponentially until $t=0.5L/c$, the time
at which scattered and reflected radiation first returns to the atom.
The amplitudes to find the other atoms excited are identically zero
until radiation first reaches them: the scattering atom first becomes
excited at $t=0.25L/c$ and the detector atom is first excited at
$t=0.5L/c$.  The three decay constants which characterize the
spontaneous emission rates of each of the atoms emerge naturally in
terms of the parameters of the Hamiltonian as
\begin{equation}
\gamma_j = \frac{\pi\vert\Omega_j\vert^2}{\Delta_{\rm c}} = 
\vert\Omega_j\vert^2\frac{L}{c}.
\label{eq:gamma:deg}
\end{equation}
The causal nature of the dynamics is evident in that all disturbances
are propagated at the speed of light $c$ via the quantum field. The
abrupt changes in the complex amplitudes at intervals of $0.5L/c$ are
a manifestation of the finite speed of light and the atomic spacing of
$0.25L$.

The abrupt changes are manifested in our analytic solution for the
complex amplitudes $c_j(t)$ and $b_m(t)$ as sums of terms with step
functions that ``turn on'' at successively later intervals of
$0.5L/c$.  In the following formulas giving these amplitudes we
truncate the expressions so that only the first excitations of atoms 2
and 3 are included.  We also note that the following equations are
specific in some details to the atomic positions $z_i$ used in this
paper.  The positions of the atoms enter via the coupling constants
$g_{jm}$, and changes in positions will result in changes in
``turn-on'' times and relative phases of terms.  One manifestation of
our large-cavity limit $\omega\gg\Delta_{\rm c}$ is that the source atom
decay at the free-space rates is clearly visible before
interruption by reflected and scattered radiation, for example at
times $t<0.5L/c$ in figure \ref{f:sol2}.  The initial atomic decay
rates given by equation (\ref{eq:gamma:deg}) are {\em not} affected by
the positions of the atoms; the effects of position dependence in the
coupling constants compensate to keep the total initial decay rates
independent of position.  (Examples of the relationship between
interrupted free-space decay and cavity-modified decay rates are given
in \cite{GIE96}.)  Later in this paper we will consider decay times
that are much shorter than $L/c$ so that the effects of all initial
excitations are complete before reflected or mutiply scattered
radition excites the atoms.  The conclusions of this paper concerning
delay times are unaffected by any details of position dependence.

The complete time dependence of the system is given by the following
set of amplitudes (in which the time is scaled in units of $L/c$):
\begin{eqnarray}
\fl c_1(t) =  \exp\left(-\frac{\gamma_1}{2}t\right) + 
              \Theta\left(t-\frac{1}{2}\right)
              \frac{\gamma_1}{(\gamma_1-\gamma_2 +\rmi 2\delta)^2} \nonumber \\
\lo{\times} \left\{
	  \exp\left[-\frac{\gamma_1}{2}\left(t-\frac{1}{2}\right)\right]
	  \left[\frac{(\gamma_1 - 2\gamma_2 + \rmi 2\delta)
			(\gamma_1 - \gamma_2 + \rmi 2\delta)}{2}
	  \left(t-\frac{1}{2}\right)- \gamma_2\right] \right. \nonumber \\
          \left. +\gamma_2\exp\left[-\left(\frac{\gamma_2}{2}-\rmi\delta\right)
          \left(t-\frac{1}{2}\right)\right]\right\} + \cdots  
\label{eq:c1:gen}  
\end{eqnarray}
\begin{equation}
\fl c_2(t) = \Theta\left(t-\frac{1}{4}\right)
      \frac{\sqrt{\gamma_1\gamma_2}}{\gamma_1- \gamma_2 + \rmi 2\delta}
      \left\{\exp\left[-\frac{\gamma_1}{2}\left(t-\frac{1}{4}\right)\right] - 
      \exp\left[-\left(\frac{\gamma_2}{2}-\rmi \delta\right)
         \left(t-\frac{1}{4}\right)\right]\right\} + \cdots
\label{eq:c2:gen}
\end{equation}
\begin{eqnarray}
\fl c_3(t) = \Theta\left(t-\frac{1}{2}\right)\sqrt{\gamma_1\gamma_3}\left\{
		\exp\left[-\frac{\gamma_1}{2}\left(t-\frac{1}{2}\right)\right]
		\frac{(\gamma_1+\rmi 2\delta)}{(\gamma_1
	-\gamma_3)(\gamma_1-\gamma_2 + \rmi 2\delta)} \right. \nonumber \\
\lo- \left. \exp\left[-\left(\frac{\gamma_2}{2}-\rmi \delta\right)
                                  \left(t-\frac{1}{2}\right)\right]
		\frac{\gamma_2}{(\gamma_1-\gamma_2+\rmi 2\delta)
				(-\gamma_2+\gamma_3+\rmi 2\delta)} 
		\right. \nonumber \\ 
\lo+ \left. \exp\left[-\frac{\gamma_3}{2}\left(t-\frac{1}{2}\right)\right]
		\frac{(\gamma_3+\rmi 2\delta)}{(-\gamma_1
	+\gamma_3)(-\gamma_2+\gamma_3 +\rmi 2\delta)} \right\}
					 + \cdots 
\label{eq:c3:gen}
\end{eqnarray}
\begin{eqnarray}
 \fl b_m(t) = \frac{\rmi 2g_{1m}}{(\gamma_1-\rmi 2m\pi)}
	\left[\exp\left(-\frac{\gamma_1}{2}t\right)-
           \exp\left(-\rmi m\pi t\right)\right] 
	+ \Theta\left(t-\frac{1}{4}\right)
        \rmi 2g_{2m}\sqrt{\gamma_1\gamma_2} \nonumber \\
\lo\times \left\{
       \frac{\exp\left[-\frac{\gamma_1}{2}\left(t-\frac{1}{4}\right)\right]}
          {(\gamma_1-\rmi 2m\pi)(\gamma_1-\gamma_2+\rmi 2\delta)}  
	 + \frac{\exp\left[-\rmi m\pi\left(t-\frac{1}{4}\right)\right]}
              {(\gamma_1 - \rmi 2m\pi)[\gamma_2 -\rmi 2(\delta+m\pi)]} -
	      \right. \nonumber \\
\lo - \left. \frac{\exp\left[-\left(\frac{\gamma_2}{2}+\rmi \delta\right)
	    \left(t-\frac{1}{4}\right)\right]}
	  {\left[\gamma_2 -\rmi 2(\delta + m\pi)\right]
			(\gamma_1 - \gamma_2 + \rmi 2\delta)}
	\right\} + \cdots 			
\label{eq:b:m:gen}
\end{eqnarray}
(We note that if we had made a different assumption about the size of 
resonant wavelength relative to the length of the cavity in our discussion
immediately preceding equation (\ref{eq:enumerate}), then the complex
phases associated with terms in the equations above would be different.)

In the following sections we will focus on two quantities: $c_3(t)$,
the amplitude to find the detector atom excited, and $\langle\hat{\cal
E}^2\rangle$ the expectation value of the square of the electric field
operator, which is proportional to the field intensity.  (The
expectation value of the field operator itself is zero for any state
with the form of equation (\ref{eq:gen:psi}).)  In our investigation of
$c_3(t)$ we will consider only the displayed term in equation
(\ref{eq:c3:gen}) describing the initial excitation of the detector atom.
Similarly, we will investigate $\langle\hat{\cal E}^2\rangle$ in
regions to the right of the scattering atom, and at times that exclude
multiple scattering effects.

It is useful to rewrite $c_3(t)$ as the sum of two pieces: the
amplitude $c^0_3(t)$ for atom 3 to be excited in the absence of the
scattering atom (or, equivalently, when $\gamma_2=0$), and
$c_3^{\rm s}(t)$, the amplitude that is attributable to scattering.  The 
total amplitude is thus
\begin{equation}
c_3(t)\equiv c_3^0(t) + c_3^{\rm s}(t).  
\label{eq:separate:c3}
\end{equation}
Setting $\gamma_2=0$ in equation (\ref{eq:c3:gen}) gives
\begin{equation}
\fl c^{0}_3(t) = \Theta\left(t-\frac{1}{2}\right)
	\frac{\sqrt{\gamma_1\gamma_3}}{\gamma_1-\gamma_3}\left\{
	\exp\left[-\frac{\gamma_1}{2}\left(t-\frac{1}{2}\right)\right] -
	\exp\left[-\frac{\gamma_3}{2}\left(t-\frac{1}{2}\right)\right]\right\},
\label{eq:c30}
\end{equation}
and subtracting this from equation (\ref{eq:c3:gen}) gives 
\begin{eqnarray}
\fl c_3^{\rm s}(t) = \Theta\left(t-\frac{1}{2}\right)
	\gamma_2\sqrt{\gamma_1\gamma_3}
	\left\{	\frac{\exp\left[-\frac{\gamma_1}{2}
                         \left(t-\frac{1}{2}\right)\right]}
			{(\gamma_1-\gamma_3)(\gamma_1-\gamma_2+\rmi 2\delta)}
		+ \frac{\exp\left[-\left(\frac{\gamma_2}{2}-\rmi \delta\right)
		    \left(t-\frac{1}{2}\right)\right]}
			{(\gamma_1-\gamma_2+\rmi 2\delta)
			(-\gamma_2+\gamma_3+\rmi 2\delta)}
				\right.		\nonumber \\
\lo + \left.
        \frac{\exp\left[-\frac{\gamma_3}{2}\left(t-\frac{1}{2}\right)\right]}
			{(-\gamma_1+\gamma_3)
	                (-\gamma_2 + \gamma_3 + \rmi 2\delta)} \right\}.
\label{eq:c3s}
\end{eqnarray}

As an alternative to finding the time-dependence of the
excitation amplitude for the detector atom, we can characterize the
transmitted field itself without recourse to the details of the
detector.  Standard photodetection theory \cite{SCU97} suggests
calculation of the expectation value $\langle \hat{\cal
E}^{(-)}(z,t)\hat{\cal E}^{(+)}(z,t)\rangle$, where $\hat{\cal
E}^{(+)}(z,t)$ and $\hat{\cal E}^{(-)}(z,t)$ correspond to the
decomposition of the interaction representation field operator into
positive and negative frequency parts.  This is equivalent to 
the calculation of the expectation value of the normally ordered
intensity operator $\langle :\hat{\cal E}^2:\rangle$. 

Using the electric field operator in the form given in reference 
\cite{MEY99}, we write the expectation value of the square of
the field as
\begin{equation}
\fl \langle :\hat{\cal{E}}^2:\rangle = \langle \psi(t) \vert :\left\{
	\sum_m \left(\frac{\hbar\omega_m}{\epsilon_0 V}\right)^{1/2}  
		\left(a_m + a_m^\dagger\right)
		\sin\left[(m_0 + m)\frac{\pi z}{L}\right] \right\}^2: 
		\vert \psi(t)\rangle.
\end{equation}
In the limit considered in this paper we can replace the frequencies
$\omega_m$ under the radical with the constant $\omega_1^{\rm (at)}$.
After expanding the state vector as in equation (\ref{eq:gen:psi}),
normally ordering the operators, and evaluating the sums, the
expectation value can be written in terms of the amplitudes $b_m(t)$
to find the photon in the various cavity modes:
\begin{equation} 
\langle :\hat{\cal{E}}^2:\rangle 
	= \left(\frac{2\hbar\omega^{\rm (at)}_1}{\epsilon_0 V}\right)
	  \left\vert \sum_m b_m(t)\sin\left[(m_0 + m)\frac{\pi z}{L}\right]
			\right\vert^2.
\label{eq:e2:expect}
\end{equation}
Evaluation of this expression gives a space- and time-dependent
representation of the localization of the energy of the photon
\cite{LIG02,BUZ99}.

The expression for $\langle :\hat{\cal{E}}^2:\rangle$ in
equation (\ref{eq:e2:expect}) is the square of a complex number that 
is analogous to the classical complex analytic signal.  We label this
quantity ${\cal E}_{\rm q.m}$, i.e.,
\begin{equation}
{\cal E}_{\rm q.m.} = \left(\frac{2\hbar\omega^{\rm (at)}_1}
		{\epsilon_0 V}\right)^{1/2}
	  \sum_m b_m(t)\sin\left[(m_0 + m)\frac{\pi z}{L}\right].
\end{equation}
The overall phase of this quantity is clearly arbitrary; in what 
follows we retain the phase that comes from a direct evaluation this 
equation.

We note that quantity ${\cal E}_{\rm q.m.}$ is very closely related to
what has been identified as ``the `electric field' associated with [a]
single photon state'' by Scully and Zubairy \cite{SCU97}.  They
consider states which are the product of separable atomic and
field states. They conclude that for single photon states $\vert
\psi_\gamma\rangle$ the quantity $\langle 0\vert
\hat{E}^{(+)}\vert\psi_\gamma \rangle$ ``can be interpreted as a kind
of a wave function for a photon.'' They also demonstrate the
classical nature of this quantity for field states produced by
spontaneous emission.  The present paper extends this kind of analysis
to define an analog to the classical analytic signal even in cases
in which field and atomic variables are entangled, and it extends the
quantum-classical correspondence to fields that include scattering
from two-level atoms.

For ease of comparison with previous results for the detector atom, 
we fix $z=3 L/4$ in what follows.  With no scattering atom present
we find (see Appendix)
\begin{equation}
{\cal E}^{0}_{\rm q.m.} = -\rmi\Theta\left(t-\frac{1}{2}\right)
		\left(\frac{\hbar\omega^{\rm (at)}_1\gamma_1\pi}
		{2\epsilon_0 V\Delta_{\rm c}}\right)^{1/2}
                  \exp\left[-\frac{\gamma_1}{2}
		  \left(t-\frac{1}{2}\right)\right].
\label{eq:analytic:sig:0}
\end{equation}
The energy density passing the point $z=3 L/4$ exhibits an abrupt
turn-on (because of the initial conditions we have chosen) followed by
exponential decay \cite{LIG02,BUZ99,SCU97}.  With a scattering atom
present at $z=L/2$ we find
\begin{eqnarray}
\fl {\cal E}_{\rm q.m.}
	= -\rmi \Theta\left(t-\frac{1}{2}\right)
	  \left(\frac{\hbar\omega^{\rm (at)}_1\gamma_1\pi}
                     {2\epsilon_0 V\Delta_{\rm c}}\right)^{1/2}
	   \left(\exp\left[-\frac{\gamma_1}{2}\left(t-\frac{1}{2}\right)\right]
           -\rmi \frac{\gamma_2}{[2\delta -\rmi (\gamma_1-\gamma_2)]} 
                   \right. \nonumber \\
        \left. 
  \times \left\{\exp\left[-\frac{\gamma_1}{2}\left(t-\frac{1}{2}\right)\right] 
   -    \exp\left[-\left(\frac{\gamma_2}{2}-\rmi \delta\right)
            \left(t-\frac{1}{2}\right)\right]\right\}\right)
\label{eq:analytic:sig}.
\end{eqnarray}
Using only the first exponential term leads to the previous result
with no scattering atom present; the effect of the scattering is
contained in the remaining terms.  Calculation of the expectation
value of the square of the field will exhibit interference between
the terms of this expression, leading to effects similar to those
noted in reference \cite{DRO00}.

\section{Comparison of classical and quantum mechanical scattering}
\label{sec:comparison}

In order to compare the quantum mechanical ``pulses'' derived in
section \ref{sec:quantum} to classical analogs it is necessary to
extend the treatment reviewed in section \ref{sec:classical} to
include the effects of transients. This can be done in the frequency
domain using Fourier techniques, or in the time domain by explicitly
including transients in the solution of the equation of motion of the
charged oscillators, and calculating the fields re-radiated by the
transients.  The quantum mechanical results of section \ref{sec:quantum} have
been derived in the time domain, and to emphasize the analogy between
the classical and quantum cases we review the classical time domain
calculation.

The quantum pulses of equation (\ref{eq:analytic:sig:0}) that are incident
on the scattering atom have the classical analog
\begin{equation} 
{\cal E}^{0}_{\rm cl.} = \Theta\left(t-\frac{1}{2}\right)
		C \exp\left[-\left(\frac{\gamma_1}{2}+\rmi\omega_1\right)
		  \left(t-\frac{1}{2}\right)\right]
\label{eq:analytic:sig:0:cl},
\end{equation}
where $C$ is a constant.  Using the field of equation
(\ref{eq:analytic:sig:0:cl}) as the driving field in the equation of
motion of a driven charged oscillator with damping constant $\gamma_2$
and resonance frequency $\omega_2$, it is straightforward to find
the motion of the oscillator.  For the initial conditions $x(0)=0$
and $\dot{x}(0)=0$, and assuming that $\omega_2\gg\gamma_2$, we have
\begin{eqnarray}
\fl x(t)\simeq \Theta\left(t-\frac{1}{2}\right)
\frac{qC/m}{(\omega_2^2-\omega_1^2+\rmi \omega_1(\gamma_1-\gamma_2)
		+ \gamma_1^2/4 - \gamma_1\gamma_2/2)}\nonumber \\
\lo \times \left\{\exp\left[-\left(\frac{\gamma_1}{2}+\rmi \omega_1\right)
               \left(t-\frac{1}{2}\right)\right]-
		\exp\left[-\left(\frac{\gamma_2}{2}+\rmi \omega_2\right)
		  \left(t-\frac{1}{2}\right)\right]\right\}.
\end{eqnarray}
Calculating the re-radiated field from the motion of the charges as 
in equation (\ref{eq:e:sc}), and making the approximations 
$\delta = \omega_1-\omega_2 \ll \omega_j$  and $\gamma_j\ll\omega$,
gives the scattered field as
\begin{eqnarray}
\fl {\cal E}^{s}_{\rm cl.} = -\rmi\Theta\left(t-\frac{1}{2}\right)
	\exp\left(-\rmi\omega_1 t\right) \frac{Cf\gamma_2}{[2\delta - 
         \rmi (\gamma_1-\gamma_2)]}        \nonumber \\
\lo \times
        \left\{\exp\left[-\frac{\gamma_1}{2}\left(t-\frac{1}{2}\right)\right]-
	\exp\left[-\left(\frac{\gamma_2}{2}-\rmi \delta\right)
	  \left(t-\frac{1}{2}\right)\right]\right\}.
\end{eqnarray}
We note that the functional form of this equation giving the classical
scattered field is identical to the portion of equation
(\ref{eq:analytic:sig}) describing the quantum scattered field.  In
addition, we note that this functional form is also reflected in
$c_3(t)$.  In the limit of rapid detector response, i.e.,
$\gamma_3\gg\gamma_1,\gamma_2$, this amplitude is
\begin{eqnarray}
\fl c_3(t) \simeq -\Theta\left(t-\frac{1}{2}\right)
	\sqrt{\frac{\gamma_1}{\gamma_3}}\left( 
		\exp\left[-\frac{\gamma_1}{2}\left(t-\frac{1}{2}\right)\right]
            -\rmi \frac{\gamma_2}{[2\delta - \rmi (\gamma_1-\gamma_2)]}
             \right.                     \nonumber \\
\lo\times \left. 
        \left\{\exp\left[-\frac{\gamma_1}{2}\left(t-\frac{1}{2}\right)\right] 
        -\exp\left[-\left(\frac{\gamma_2}{2}-\rmi \delta\right)
         \left(t-\frac{1}{2}\right)\right]\right\}\right),
\end{eqnarray}
and the contribution due to scattering is again the same.  (Our use of 
a ``detector'' which is an atom identical to the ``source,'' except with 
a more rapid decay rate, is not meant to correspond to any real experiment.
It is an idealization meant to illustrate the effects of the quantum 
field on a simple system with a fast response.) 

Thus we see the exact functional form of the classical field reflected
in two quantum mechanical quantities: the probability amplitude of the
detector atom $c_3(t)$, and in our quantum field ``amplitude'' ${\cal
E}_{\rm q.m.}$.  The classical and quantum expressions all include a
``steady-state'' term (which drops off at the slow decay rate of source 
atoms, $\gamma_1/2$).  In the limit $\gamma_1\ll\gamma_2$ the radiation
emitted by the source has a very narrow line-width compared to the
scattering atom, and this steady-state term simplifies to the form of
equation (\ref{eq:e:tr:approx}) given by the simple classical theory
of section \ref{sec:classical}.

The expression for the quantum field ``amplitude'' ${\cal E}_{\rm
q.m.}$ contains a time-dependent phase that is identical to the phase
of the classical field, and the probability amplitude $c_3(t)$
contains terms due to incident and scattered fields with relative
phases that match classical expectations.  The concept of the phase of
a quantum field has been the subject of investigation from the early
days of quantum mechanics \cite{ORS00,PER82}.  We note, however, that
most previous work on the phase of a quantum field has focused on
defining a meaningful and mathematically well-behaved phase operator
for linear combinations of multiply occupied states of a single-mode
field; in this paper we see manifestations of classical phases in a
multimode field which contains a single photon.
 
In the case of classical scattering considered in section
\ref{sec:classical}, the effect of a single scattering event is
considered to be small, and exponential phase rotation and attenuation
of the transmitted field are the result of many scattering events.  In
our quantum model the magnitude of the scattering is determined by
$\Omega_2$ (or equivalently $\gamma_2$) which characterizes the
coupling of atom 2 to the field.  In our one-dimensional model the
coupling to the incident field and the decay rate of atom 2 are both
completely determined by the single parameter $\Omega_2$, which means
that it is not possible to make the effect of the scattering small
without simultaneously making the line-width of atom 2 very narrow.
In a fully three-dimensional model the decay rate of atom 2 would be
the result of the atom's coupling to many more modes, and not just
those containing the incident field.  This scattering into other modes
would reduce the scattering in the forward direction (the direction of
the detector) from the amount predicted in our simple model.  Of
course in a fully-three dimensional model there are other effects
also: the fraction of source radiation emitted in the direction of the
scatterer (and detector) would be reduced, and the scatterer would be
driven by a weaker field and would emit less total radiation. The net
effect is that the excitation of the detector due to the scattered
field will be reduced relative to the direct excitation.  If the
relative effect of forward scattering is reduced by a factor $f$, then
a more realistic expression for the excitation of the detector atom is
\begin{equation}
c_3(t) = {\cal N}\left(c_3^0(t) + f c_3^{\rm s}(t)\right),
\label{eq:f}
\end{equation}
where ${\cal N}$ gives an overall reduction in the excitation.  We
have used the same symbol $f$ here that we used earlier for the
dimensionless parameter which characterizes the magnitude of classical
scattering in section \ref{sec:classical}.  The quantum superposition
of equation (\ref{eq:f}) is analogous to the classical field
superposition of equations (\ref{eq:e:tr}) and (\ref{eq:e:tr:approx}).

\section{Temporal-center-of-gravity delay}
\label{sec:cog:delay}

The group delay of a classical pulse has a clear interpretation for
quasi-monochromatic pulses: it is the delay in the arrival of the peak
of a pulse compared to the time expected for propagation through a
vacuum.  The pulses investigated in this paper have sharp leading
edges, and this lack of a smooth modulation envelope means that the
results of simple classical theory for quasi-monochromatic pulses
should not be expected to be a sufficient guide to full understanding.
In this section we investigate ``temporal-center-of-gravity''
delays in several classical and quantum mechanical quantities.

\begin{figure}
\begin{center}
\includegraphics{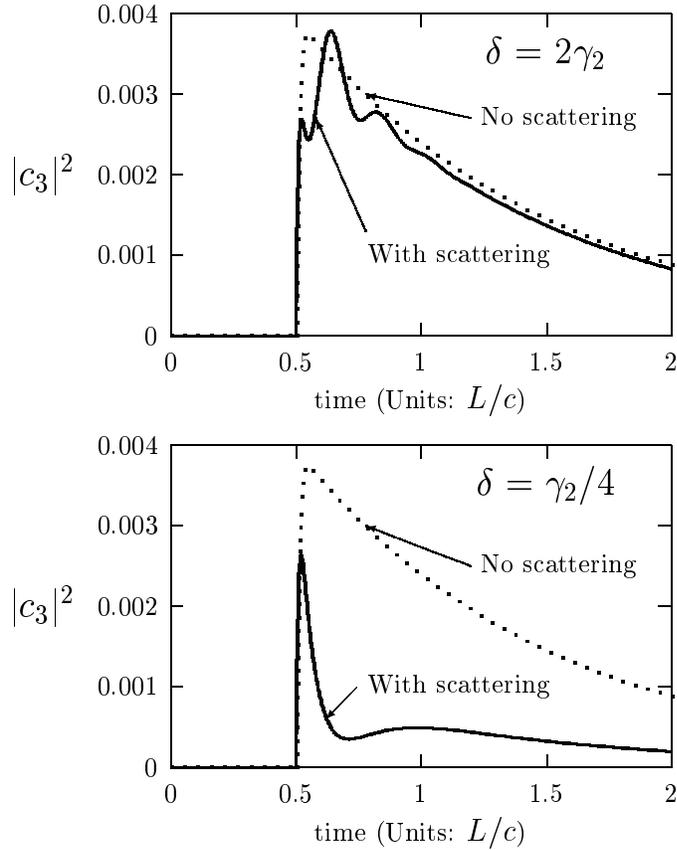}
\end{center}
\caption{Probability for the detector atom to be excited as a function
of time.  The top figure is for a detuning $\delta$ in the region of
normal dispersion, and the bottom is in the region of anomalous
dispersion.  The effect of the ``pulse re-shaping'' is different in
the two graphs.  In the top graph probability is preferentially
removed from earlier times, while the converse is true in the lower
graph.  The decay rates of the atoms are $\gamma_1 = 1$, $\gamma_2 =
16$, and $\gamma_3 = 256$ (in the units of the figure).}
\label{f:c3}
\end{figure}

The first delay we investigate is derived from $c_3(t)$, the amplitude
for the detector atom to be excited.  As we have argued previously,
this amplitude will reflect the strength of the incident field in the
limit that the response time of this atom is very small compared with
other time scales, i.e., $\gamma_3\gg \gamma_1, \gamma_2$.  The effect
of the scattering on this amplitude is evident in figure \ref{f:c3},
in which $\vert c_3(t)\vert^2$ is plotted for two values of the
detuning $\delta$, and also for the case in which no scattering atom
is present.  (For clarity in this figure we have only included the 
terms describing the initial ``turn-on'' of excitation; we have 
not included effects due to reflection and multiple scattering.)
For all detunings, $c_3(t)=0$ for all times earlier than $t=0.5L/c$, as is 
expected; all effects on the detector atom occur at times that preserve
causality.  The qualitative shapes of the detector response depend
critically on the detuning $\delta$.  For detunings in the region of 
normal dispersion the response is relatively reduced at early times
compared to later times, resulting in a qualitative delay in detection
of the photon.  For detunings in the region of anomalous dispersion 
($-\gamma_2/2 <\delta <\gamma_2/2$) the converse is true.  

We quantify these ideas by identifying an effective arrival time of
the photon with the temporal center of gravity of the probability that
the detector atom is excited, i.e.,
\begin{equation}
t_{\rm arrival} =  \frac{\int t\vert c_3(t)\vert^2\, \rmd t}
		{\int \vert c_3(t)\vert^2\, \rmd t}.
\end{equation}
(In evaluating the integrals in this equation, we use only the term in
the series of equation (\ref{eq:c3:gen}) that ``turns on'' at $t=1/2$,
and assume that the decay rates and distances are such that the effect
of multiple scattering is negligible. These decay rates are very much 
faster than those displayed in the figures; the rates in the figures 
were chosen so that the the atomic excitation dynamics and the causal 
delays occurred on the same time scale, and could be illustrated on the 
same graphs.)  The delay imposed by the medium is then just the 
difference in the arrival times calculated with and without a 
scattering atom present, 
\begin{equation}
\Delta t_{c_3} = \frac{\int t \vert c_3(t)\vert^2\, \rmd t}
			{\int \vert c_3(t)\vert^2\, \rmd t}- 
		  \frac{\int t \vert c_3^0(t)\vert^2\, \rmd t}
			{\int  \vert c_3^0(t)\vert^2\, \rmd t}.
\end{equation}
To explore the effect of weak scattering we rewrite $c_3(t)$ in 
the form of equation (\ref{eq:f}), and assume that $f\ll 1$.  Our quantum
mechanical delay becomes, to first order in $f$,
\begin{eqnarray}
\Delta t_{c_3} &=& \frac{\int t\vert c_3^0(t)+ fc_3^{\rm s}(t)\vert^2\, \rmd t}
			{\int \vert c_3^0(t)+ fc_3^{\rm s}(t)\vert^2\, \rmd t}
- \frac{\int t\vert c_3^0(t)\vert^2\, \rmd t}
	{\int \vert c_3^0(t)\vert^2\, \rmd t} \nonumber \\
	&\simeq& 2f \left[\frac{\int t\, 
	        \mbox{Re}\left[c_3^0(t) c_3^{\rm s}(t)^\ast\right]\, \rmd t}
		{\int  \vert c_3^0(t)\vert^2 \rmd t} 
                   - \frac{\int t\vert c_3^0(t)\vert^2\, \rmd t 
	    \int \mbox{Re}\left[c_3^0(t) c_3^{\rm s}(t)^\ast\right]\, \rmd t}
			{\left(\int \vert c_3^0(t)\vert^2\, \rmd t\right)^2}
			\right].
\label{eq:delta_t_qm}
\end{eqnarray}
It is straightforward to evaluate the integrals in equation
(\ref{eq:delta_t_qm}) using the expressions for $c_3^0(t)$ and
$c_3^{\rm s}(t)$ from equations (\ref{eq:c30}) and (\ref{eq:c3s}).
After taking the limit $\gamma_3\rightarrow\infty$ and then letting
$\gamma_1\rightarrow 0$ we arrive at the following expression for the
quantum time delay:
\begin{equation}
\Delta t_{c_3} = 4f\gamma_2\frac{(4\delta^2-\gamma_2^2)}{(4\delta^2 + 
     \gamma_2^2)^2}.
\label{eq:group:delay:qm}
\end{equation}
Comparing this to equation (\ref{eq:group:delay}) shows that
``temporal-center-of-gravity'' delay time for this specific pulse is
identical in functional form to the classical group delay.  The
magnitude of the ``temporal-center-of-gravity'' delay is, however,
twice that given by the group delay.  (The
``temporal-center-of-gravity'' delay for classical pulses with the
form given in equation (\ref{eq:analytic:sig:0:cl}) is also twice as
large as the group delay.)

The delay in the arrival time that we have defined is the result of 
the reshaping of the ``pulse'' of excitation of the detector atom.
The effect of scattering in the region of anomalous dispersion 
is to reduce preferentially the probability that the detector
atom will be excited at large times; a short spike of probability
at early times remains, shifting the ``center-of-gravity'' of the 
excitation to earlier times.  Despite the appearance of negative 
delays, no excitation of the detector atom occurs at times $t<0.5L/c$ 
for any values of the parameters in our model.  Although we have
not demonstrated it explicitly in this work, we are confident that 
delays of quasi-monochromatic quantum pulses can be explained in the 
same manner.  

Because ${\cal E}_{\rm cl.}$, ${\cal E}_{\rm q.m.}$, and
$c_3(t)$ all have the same functional form (in the large $\gamma_3$
limit) it is easy to see that equivalent delays can be derived from
the classical field using equation (\ref{eq:cl:tcom:delay}), or from
the quantum field using the analogous equation
\begin{equation}
\fl \Delta t_{\langle{\cal E}^2\rangle}=
		\left(\frac{\int t \langle:\hat{{\cal E}}(z=3L/4)^2:\rangle\,
 					\rmd t}
                {\int \langle :\hat{{\cal E}}(z=3L/4)^2:\rangle\, \rmd t}
		\right)_{\rm with\, scatterer}
  -\left(\frac{\int t \langle :\hat{{\cal E}}(z=3L/4)^2:\rangle\,\rmd t}
                {\int \langle :\hat{{\cal E}}(z=3L/4)^2:\rangle\, \rmd t}
                \right)_{\rm no\, scatterer}. 
\end{equation}

\section{Conclusion}
We have considered simple classical and quantum models of propagation
of light through dispersive media, and we have identified quantum
mechanical quantities that have exact analogs in the classical
scattered field.  Scattering induces delays in the quantum model which
are identical to the delays of a classical field.  Because we use
standard field-theoretic techniques of quantum optics to account for
the creation and absorption of photons in our model, we have a
well-defined initial condition in which all of the energy is localized
at the position of the initially excited atom.  This enables us to
demonstrate that no effects propagate faster than the vacuum speed of
light $c$, in spite of the appearance of negative delays that
seemingly correspond to superluminal group velocities.  In a companion
paper we apply similar techniques to study delays of photons in 
media exhibiting electromagnetically induced transparency \cite{PUR02b}.

\appendix 

\section{Solution using Laplace Transforms} 
Taking the Laplace transform of the coupled differential equations 
(\ref{eq:c:schro})-(\ref{eq:b:schro}) gives the coupled algebraic 
equations
\begin{equation}
\rmi \left(s\tilde{c}_1(s) - 1\right) = \sum_m g_{1m}\tilde{b}_m(s), 
\label{eq:lt1}
\end{equation}
\begin{equation}
\rmi s\tilde{c}_2(s) = -\delta\tilde{c}_2(s) + \sum_m g_{2m}\tilde{b}_m(s), 
\label{eq:lt2}
\end{equation}
\begin{equation}
\rmi s\tilde{c}_3(s) = \sum_m g_{3m}\tilde{b}_m(s), 
\label{eq:lt3}
\end{equation}
\begin{equation}
\rmi s\tilde{b}_m(s) = m\Delta_{\rm c} \tilde{b}_m(s) + 
                         \sum_j \tilde{c}_j(s)g^\ast_{jm}. 
\label{eq:lt4}
\end{equation}

The idea behind our solution is straightforward: solve these algebraic
equations for the quantities $\tilde{c}_j(s)$ and $\tilde{b}_m(s)$ and
then perform an inverse Laplace transform to recover the time
dependence of $c_j(t)$ and $b_m(t)$.  The details of carrying out such
calculations are quite involved, and were completed with the aid of a
computer algebra system.\footnote{Mathematica notebooks used to
perform the calculations are available from the authors.}  In this
appendix we outline our approach and present some of our intermediate
results.

We begin by solving equation (\ref{eq:lt4}) for $\tilde{b}_m(s)$, and
substitute the result in the first three equations, giving
\begin{equation}
s\tilde{c}_1 -1 =-\rmi\Delta_{\rm c}\left(f_{11}\tilde{c}_1+f_{12}\tilde{c}_2 +
                   f_{13}\tilde{c}_3\right), \label{eq:ltp1}
\end{equation}
\begin{equation}
(s - \rmi \delta)\tilde{c}_2 = -\rmi\Delta_{\rm c}\left(f_{21}\tilde{c}_1 + 
                  f_{22}\tilde{c}_2 +f_{23}\tilde{c}_3\right),\label{eq:ltp2}
\end{equation}
\begin{equation}
s \tilde{c}_3 = -\rmi \Delta_{\rm c}\left(f_{31}\tilde{c}_1 + 
                  f_{32}\tilde{c}_2 + f_{33}\tilde{c}_3\right),
				\label{eq:ltp3} 
\end{equation}
in which we have defined the dimensionless sums
\begin{equation}
f_{ln} = \frac{1}{\Delta_{\rm c}^2}\sum_m\frac{g_{lm}g_{nm}^\ast}
	{\frac{\rmi s}{\Delta_{\rm c}} - m}.
\end{equation}

In the limit in which the atomic resonance frequencies are very much
greater than the fundamental frequency of the cavity, i.e.,
$\omega_j^{\rm (at)}\gg\Delta_{\rm c}$, these sums may be approximated
by extending the range for $m$ from $-\infty$ to $+\infty$, in which
case the sums have relatively simple representations in terms of
trigonometric functions.  An explicit example of one of these sums (for 
atoms at positions $z_1=L/4$, $z_2=L/2$, and $z_3=3L/4$) is 
\begin{eqnarray}
f_{11} &\simeq \frac{\gamma_1}{8\Delta_{\rm c}}\left\{
     \cot\left[\frac{\pi}{4}\left(\frac{\rmi s}{\Delta_{\rm c}}-1\right)\right]
   +2\cot\left[\frac{\pi}{4}\left(\frac{\rmi s}{\Delta_{\rm c}}-2\right)\right]
    +\cot\left[\frac{\pi}{4}\left(\frac{\rmi s}{\Delta_{\rm c}}-3\right)\right]
        \right\}  \nonumber \\
	&= -\rmi \left(\frac{\gamma_1}{4\Delta_{\rm c}}\right)
		\frac{\sinh\frac{3\pi s}{4\Delta_{\rm c}}}
		{\cosh\frac{\pi s}{4\Delta_{\rm c}}\cosh\frac{\pi s}
                           {2 \Delta_{\rm c}}}.
\end{eqnarray}

After solving equations (\ref{eq:ltp1})-(\ref{eq:ltp3}) for the
quantities $\tilde{c}_j(s)$ in terms of the sums $f_{jm}$, we rewrite
the hyperbolic trigonometric functions resulting from the sums in
terms of exponential functions; we also let $c/L=1$ at this point in
the calculation. We then expand the resulting expressions in powers of
$\exp(-s/4)$, and the time dependence of the system is recovered by a
term-by-term inverse Laplace transform of the expansion.  The step
function turn-on of the resulting time dependence arises because of
the factors $\exp(-ns/4)$ in the expansion.  The lowest order terms in
our expansions of the Laplace transforms are given here:
\begin{eqnarray}
\tilde{c}_1(s) = \frac{2}{2s + \gamma_1} + 
                \frac{4\exp(-s/2)\gamma_1(s+\gamma_2 -\rmi \delta)}
		{(2s+\gamma_1)^2(2s+\gamma_2-\rmi 2\delta)} + \cdots\\
\tilde{c}_2(s) = -\frac{2\exp(-s/4)\sqrt{\gamma_1\gamma_2}}
		   {(2s+\gamma_1)(2s+ \gamma_2-\rmi 2\delta)} 
				+ \cdots \\
\tilde{c}_3(s) = -\frac{4\exp(-s/2)\sqrt{\gamma_1\gamma_3}(s -\rmi \delta)}
			{(2s + \gamma_1)(2s+\gamma_3)
			(2s + \gamma_2 -\rmi 2\delta)} + \cdots\\
\tilde{b}_m(s) = \frac{2\rmi }{(s+\rmi m\pi)(2s + \gamma_1)}
		\left[\frac{\exp(-s/4)\sqrt{\gamma_1\gamma_2}g_{2m}}
				{(2s+\gamma_2 - \rmi 2\delta)} - g_{1m}\right].
\end{eqnarray}
The inverse Laplace transform of these expressions gives equations 
(\ref{eq:c1:gen})--(\ref{eq:b:m:gen}).

To calculate ${\cal E}_{\rm q.m.}$ of equation (\ref{eq:analytic:sig})
we repeat the process with the detector atom at $z=3L/4$ removed,
yielding
\begin{eqnarray}
\tilde{c}_1(s) = \frac{2}{2s + \gamma_1} + \cdots \\
\tilde{c}_2(s) = -\frac{2\exp(-s/4)\sqrt{\gamma_1\gamma_2}}
		   {(2s+\gamma_1)(2s+ \gamma_2-\rmi 2\delta)} 
				+ \cdots   \\
\tilde{b}_m(s) =\frac{2}{(\rmi s - m\Delta_{\rm c})(2s+\gamma_1)}
	           \left(g_{1m}^\ast - g_{2m}^\ast\frac{
		  \sqrt{\gamma_1\gamma_2}\exp(-s/4)}
			{(2s+\gamma_2-\rmi 2\delta)}\right) + \cdots .
\end{eqnarray}
Using the same techniques as above,  the Laplace transform of 
${\cal E}_{\rm q.m.}$ can be written as 
\begin{eqnarray}
\tilde{\cal E}_{\rm q.m.}&=\sqrt{\frac{2\hbar\omega_1^{\rm (at)}}{\epsilon_0V}}
			\sum_m \tilde{b}_m(s) \sin\left(\frac{m3\pi}{4}\right)
			\nonumber \\
	  &\rightarrow -\rmi \sqrt{\frac{2\hbar\omega_1^{\rm (at)}\gamma_1}
                                 {\pi\Delta_{\rm c}\epsilon_0V}}
		\frac{\exp(-s/2)}{(2s+\gamma_1)}\left[
		1-\frac{\gamma_2}{(2s + \gamma_2 - \rmi  2\delta)}\right]. 
			\label{eq:lt:field}
\end{eqnarray} 
The inverse Laplace transform of equation
(\ref{eq:lt:field}) gives equation (\ref{eq:analytic:sig}).

\ack
The authors thank James Supplee for helpful discussions and careful
reading of the manuscript.  One of us (T.P.) 
acknowledges support from National Science Foundation Research
Experiences for Undergraduates Program (Grant Number PHY-0097424).


\section*{References}

\begin{thebibliography}{28}
\expandafter\ifx\csname natexlab\endcsname\relax\def\natexlab#1{#1}\fi
\expandafter\ifx\csname bibnamefont\endcsname\relax
  \def\bibnamefont#1{#1}\fi
\expandafter\ifx\csname bibfnamefont\endcsname\relax
  \def\bibfnamefont#1{#1}\fi
\expandafter\ifx\csname citenamefont\endcsname\relax
  \def\citenamefont#1{#1}\fi
\expandafter\ifx\csname url\endcsname\relax
  \def\url#1{\texttt{#1}}\fi
\expandafter\ifx\csname urlprefix\endcsname\relax\def\urlprefix{URL }\fi
\providecommand{\bibinfo}[2]{#2}
\providecommand{\eprint}[2][]{\url{#2}}

\bibitem{BRI60}
\bibinfo{author}{\bibfnamefont{L.}~\bibnamefont{Brillouin}},
  \emph{\bibinfo{title}{Wave Propagation and Group Velocity}}
  (\bibinfo{publisher}{Academic}, \bibinfo{address}{New York},
  \bibinfo{year}{1960}).

\bibitem{HAU99}
\bibinfo{author}{\bibfnamefont{L.~V.} \bibnamefont{Hau}},
  \bibinfo{author}{\bibfnamefont{S.~E.} \bibnamefont{Harris}},
  \bibinfo{author}{\bibfnamefont{Z.}~\bibnamefont{Dutton}}, \bibnamefont{and}
  \bibinfo{author}{\bibfnamefont{C.~H.} \bibnamefont{Behroozi}},
  \bibinfo{journal}{Nature} \textbf{\bibinfo{volume}{397}},
  \bibinfo{pages}{594} (\bibinfo{year}{1999}).

\bibitem{KAS99}
\bibinfo{author}{\bibfnamefont{M.~M.} \bibnamefont{Kash}},
  \bibinfo{author}{\bibfnamefont{V.~A.} \bibnamefont{Sautenkov}},
  \bibinfo{author}{\bibfnamefont{A.~S.} \bibnamefont{Zibrov}},
  \bibinfo{author}{\bibfnamefont{L.}~\bibnamefont{Hollberg}},
  \bibinfo{author}{\bibfnamefont{G.~R.} \bibnamefont{Welch}},
  \bibinfo{author}{\bibfnamefont{M.~D.} \bibnamefont{Lukin}},
  \bibinfo{author}{\bibfnamefont{Y.}~\bibnamefont{Rostovtsev}},
  \bibinfo{author}{\bibfnamefont{E.~S.} \bibnamefont{Fry}}, \bibnamefont{and}
  \bibinfo{author}{\bibfnamefont{M.~O.} \bibnamefont{Scully}},
  \bibinfo{journal}{Phys. Rev. Lett.} \textbf{\bibinfo{volume}{82}},
  \bibinfo{pages}{5229} (\bibinfo{year}{1999}).

\bibitem{STE93}
\bibinfo{author}{\bibfnamefont{A.~M.} \bibnamefont{Steinberg}},
  \bibinfo{author}{\bibfnamefont{P.~G.} \bibnamefont{Kwiat}}, \bibnamefont{and}
  \bibinfo{author}{\bibfnamefont{R.~Y.} \bibnamefont{Chiao}},
  \bibinfo{journal}{Phys. Rev. Lett.} \textbf{\bibinfo{volume}{71}},
  \bibinfo{pages}{708} (\bibinfo{year}{1993}).

\bibitem{WAN00}
\bibinfo{author}{\bibfnamefont{L.~J.} \bibnamefont{Wang}},
  \bibinfo{author}{\bibfnamefont{A.}~\bibnamefont{Kuzmich}}, \bibnamefont{and}
  \bibinfo{author}{\bibfnamefont{A.}~\bibnamefont{Dogariu}},
  \bibinfo{journal}{Nature} \textbf{\bibinfo{volume}{406}},
  \bibinfo{pages}{277} (\bibinfo{year}{2000}).

\bibitem{CHI97}
\bibinfo{author}{\bibfnamefont{R.}~\bibnamefont{Chiao}} \bibnamefont{and}
  \bibinfo{author}{\bibfnamefont{A.}~\bibnamefont{Steinberg}}, in
  \emph{\bibinfo{booktitle}{Progress in Optics}}, edited by
  \bibinfo{editor}{\bibfnamefont{E.}~\bibnamefont{Wolf}}
  (\bibinfo{publisher}{Elsevier}, \bibinfo{year}{1997}),
  vol.~\bibinfo{volume}{37}, pp. \bibinfo{pages}{345--405}.

\bibitem{FEY63}
\bibinfo{author}{\bibfnamefont{R.~P.} \bibnamefont{Feynman}},
  \bibinfo{author}{\bibfnamefont{R.~B.} \bibnamefont{Leighton}},
  \bibnamefont{and} \bibinfo{author}{\bibfnamefont{M.}~\bibnamefont{Sands}},
  \emph{\bibinfo{title}{The Feynman Lectures on Physics}}
  (\bibinfo{publisher}{Addison-Wesley}, \bibinfo{address}{Reading, MA},
  \bibinfo{year}{1963}), vol.~\bibinfo{volume}{I}, chap. \bibinfo{chapter}{31
  and 32}.

\bibitem{SHE96}
\bibinfo{author}{\bibfnamefont{B.~A.} \bibnamefont{Sherwood}},
  \bibinfo{journal}{Am. J. Phys.} \textbf{\bibinfo{volume}{64}},
  \bibinfo{pages}{840} (\bibinfo{year}{1996}).

\bibitem{MIL96}
\bibinfo{author}{\bibfnamefont{P.~W.} \bibnamefont{Milonni}},
  \bibinfo{journal}{Am. J. Phys.} \textbf{\bibinfo{volume}{64}},
  \bibinfo{pages}{842} (\bibinfo{year}{1996}).

\bibitem{GRI99}
\bibinfo{author}{\bibfnamefont{D.~J.} \bibnamefont{Griffiths}},
  \emph{\bibinfo{title}{Introduction to Electrodynamics}}
  (\bibinfo{publisher}{Prentice Hall}, \bibinfo{address}{Upper Saddle River,
  NJ}, \bibinfo{year}{1999}), chap.~\bibinfo{chapter}{9},
  \bibinfo{edition}{3rd} ed.

\bibitem{TAY01}
\bibinfo{author}{\bibfnamefont{D.~F.} \bibnamefont{Taylor}},
  \bibinfo{type}{Undergraduate Honors Thesis}, \bibinfo{institution}{Bucknell
  University} (\bibinfo{year}{2001}).

\bibitem{LIG01a}
\bibinfo{author}{\bibfnamefont{M.}~\bibnamefont{Ligare}} \bibnamefont{and}
  \bibinfo{author}{\bibfnamefont{D.~F.} \bibnamefont{Taylor}}
  (\bibinfo{year}{2001}), \bibinfo{note}{paper presented at the Eighth
  Rochester Conference on Coherence and Quantum Optics},
  \urlprefix\url{http://www.eg.bucknell.edu/physics/ligare.html/}.

\bibitem{LIG02}
\bibinfo{author}{\bibfnamefont{M.}~\bibnamefont{Ligare}} \bibnamefont{and}
  \bibinfo{author}{\bibfnamefont{R.}~\bibnamefont{Oliveri}},
  \bibinfo{journal}{Am. J. Phys.} \textbf{\bibinfo{volume}{70}},
  \bibinfo{pages}{58} (\bibinfo{year}{2002}).

\bibitem{BUZ99}
\bibinfo{author}{\bibfnamefont{V.}~\bibnamefont{Bu\v{z}ek}},
  \bibinfo{author}{\bibfnamefont{G.}~\bibnamefont{Drobn\'{y}}},
  \bibinfo{author}{\bibfnamefont{M.~G.} \bibnamefont{Kim}},
  \bibinfo{author}{\bibfnamefont{M.}~\bibnamefont{Havukainen}},
  \bibnamefont{and} \bibinfo{author}{\bibfnamefont{P.~L.}
  \bibnamefont{Knight}}, \bibinfo{journal}{Phys. Rev. A}
  \textbf{\bibinfo{volume}{60}}, \bibinfo{pages}{582} (\bibinfo{year}{1999}).

\bibitem{DRO00}
\bibinfo{author}{\bibfnamefont{G.}~\bibnamefont{Drobn\'{y}}},
  \bibinfo{author}{\bibfnamefont{M.}~\bibnamefont{Havukainen}},
  \bibnamefont{and}
  \bibinfo{author}{\bibfnamefont{V.}~\bibnamefont{Bu\v{z}ek}},
  \bibinfo{journal}{J. Mod. Optics} \textbf{\bibinfo{volume}{47}},
  \bibinfo{pages}{851} (\bibinfo{year}{2000}).

\bibitem{SMI70}
\bibinfo{author}{\bibfnamefont{R.~L.} \bibnamefont{Smith}},
  \bibinfo{journal}{Am. J. Phys.} \textbf{\bibinfo{volume}{38}},
  \bibinfo{pages}{978} (\bibinfo{year}{1970}).

\bibitem{BLO77}
\bibinfo{author}{\bibfnamefont{S.~C.} \bibnamefont{Bloch}},
  \bibinfo{journal}{Am. J. Phys.} \textbf{\bibinfo{volume}{45}},
  \bibinfo{pages}{538} (\bibinfo{year}{1977}).

\bibitem{MEY99}
\bibinfo{author}{\bibfnamefont{P.}~\bibnamefont{Meystre}} \bibnamefont{and}
  \bibinfo{author}{\bibfnamefont{M.}~\bibnamefont{Sargent}},
  \emph{\bibinfo{title}{Elements of Quantum Optics}}
  (\bibinfo{publisher}{Springer}, \bibinfo{address}{Berlin},
  \bibinfo{year}{1999}), \bibinfo{edition}{3rd} ed.

\bibitem{SAR74}
\bibinfo{author}{\bibfnamefont{M.}~\bibnamefont{Sargent}},
  \bibinfo{author}{\bibfnamefont{M.~O.} \bibnamefont{Scully}},
  \bibnamefont{and} \bibinfo{author}{\bibfnamefont{W.~E.} \bibnamefont{Lamb}},
  \emph{\bibinfo{title}{Laser Physics}} (\bibinfo{publisher}{Addison-Wesley},
  \bibinfo{address}{Reading, MA}, \bibinfo{year}{1974}).

\bibitem{LOU83}
\bibinfo{author}{\bibfnamefont{R.}~\bibnamefont{Loudon}},
  \emph{\bibinfo{title}{The Quantum Theory of Light}}
  (\bibinfo{publisher}{Oxford U.~P.}, \bibinfo{address}{Oxford},
  \bibinfo{year}{1983}), \bibinfo{edition}{2nd} ed.

\bibitem{STE72}
\bibinfo{author}{\bibfnamefont{G.~C.} \bibnamefont{Stey}} \bibnamefont{and}
  \bibinfo{author}{\bibfnamefont{R.~W.} \bibnamefont{Gibberd}},
  \bibinfo{journal}{Physica} \textbf{\bibinfo{volume}{60}}, \bibinfo{pages}{1}
  (\bibinfo{year}{1972}).

\bibitem{MIL74}
\bibinfo{author}{\bibfnamefont{P.~W.} \bibnamefont{Milonni}} \bibnamefont{and}
  \bibinfo{author}{\bibfnamefont{P.~L.} \bibnamefont{Knight}},
  \bibinfo{journal}{Phys. Rev. A} \textbf{\bibinfo{volume}{10}},
  \bibinfo{pages}{1096} (\bibinfo{year}{1974}).

\bibitem{MIL75}
\bibinfo{author}{\bibfnamefont{P.~W.} \bibnamefont{Milonni}} \bibnamefont{and}
  \bibinfo{author}{\bibfnamefont{P.~L.} \bibnamefont{Knight}},
  \bibinfo{journal}{Phys. Rev. A} \textbf{\bibinfo{volume}{11}},
  \bibinfo{pages}{1090} (\bibinfo{year}{1975}).

\bibitem{GIE96}
\bibinfo{author}{\bibfnamefont{H.}~\bibnamefont{Gie{\ss}en}},
  \bibinfo{author}{\bibfnamefont{J.~D.} \bibnamefont{Berger}},
  \bibinfo{author}{\bibfnamefont{G.}~\bibnamefont{Mohs}}, \bibnamefont{and}
  \bibinfo{author}{\bibfnamefont{P.}~\bibnamefont{Meystre}},
  \bibinfo{journal}{Phys. Rev. A} \textbf{\bibinfo{volume}{53}},
  \bibinfo{pages}{2816} (\bibinfo{year}{1996}).

\bibitem{SCU97}
\bibinfo{author}{\bibfnamefont{M.~O.} \bibnamefont{Scully}} \bibnamefont{and}
  \bibinfo{author}{\bibfnamefont{M.~S.} \bibnamefont{Zubairy}},
  \emph{\bibinfo{title}{Quantum Optics}} (\bibinfo{publisher}{Cambridge
  University Press}, \bibinfo{address}{Cambridge}, \bibinfo{year}{1997}).

\bibitem{ORS00}
\bibinfo{author}{\bibfnamefont{M.}~\bibnamefont{Orszag}},
  \emph{\bibinfo{title}{Quantum Optics}} (\bibinfo{publisher}{Springer-Verlag},
  \bibinfo{address}{Berlin}, \bibinfo{year}{2000}),
  chap.~\bibinfo{chapter}{15}, \bibinfo{note}{and references therein}.

\bibitem{PER82}
\bibinfo{author}{\bibfnamefont{V.}~\bibnamefont{Pe\u{r}inov\'{a}}},
  \bibinfo{author}{\bibfnamefont{A.}~\bibnamefont{Luk\u{s}}}, \bibnamefont{and}
  \bibinfo{author}{\bibfnamefont{J.}~\bibnamefont{Pe\u{r}ina}},
  \emph{\bibinfo{title}{Phase in Optics}} (\bibinfo{publisher}{World
  Scientific}, \bibinfo{address}{Singapore}, \bibinfo{year}{1982}),
  chap.~\bibinfo{chapter}{4}, \bibinfo{note}{and references therein}.

\bibitem{PUR02b}
\bibinfo{author}{\bibfnamefont{T.}~\bibnamefont{Purdy}} \bibnamefont{and}
  \bibinfo{author}{\bibfnamefont{M.}~\bibnamefont{Ligare}},
  \bibinfo{note}{submitted to \JOB, quant-ph/0204173}.

\end{thebibliography}

\end{document}